\documentclass[doublespacing,eqsecnum]{elsart}

\usepackage{graphicx}

\usepackage{amssymb}

\begin{document}

\begin{frontmatter}

\title{Limitations on the Utility of Exact Master
Equations}
\author{G. W. Ford\thanksref{ford}}
\thanks[ford]{Permanent address: Department of Physics, University of
Michigan, Ann Arbor, MI 48109-1120},
\author{R. F. O'Connell\thanksref{roc}\corauthref{auth1}}
\corauth[auth1]{Corresponding author}
\thanks[roc]{Permanent address: Department of Physics and Astronomy,
Louisiana State University}
\ead{oconnell@phys.lsu.edu}
\address{School of Theoretical Physic s Dublin
Institute for Advanced Studies 10 Burlington Road
Dublin 4, Ireland}

\begin{abstract}
The low temperature solution of the exact master equation
for an oscillator coupled to a linear passive heat bath is known to give
rise to serious divergences.  We now show that, even in the high
temperature regime, problems also exist, notably the fact that the
density matrix is not necessarily positive.
\end{abstract}

\end{frontmatter}

\section{Introduction}

In a previous publication \cite{ford01d}, we presented a general solution
of the exact master equation for an oscillator coupled to a linear
passive heat bath.  This was achieved by solving the generalized quantum
Langevin equation for the initial value problem \cite{ford3} which
enabled us to obtain explicit expressions for the time-dependent
coefficients occuring in the exact master equation.  Mext, we
illustrated the general solution with the construction of explicit
expressions for the probability distributions for two examples: wave
packet spreading and "Schr\"{o}dinger cat" state for a free particle.
This enabled us to discover that the low temperature solution gave rise
to divergences.  The purpose of the present paper is to show that problems
also exist in the high temperature regime.  Thus, we are motivated to
extend our previous analysis to obtain explicit expressions for the
Wigner function and the elements of the density matrix, and with the aid
of these results to critically examine the equation and its solution.

At the outset we must point out that the exact master equation has
drawbacks that seriously limit its utility. For the most part these are
related to the fact that the initial state is necessarily one in which
the particle and heat bath are uncoupled, while the subsequent motion is
described by the fully coupled system. The most serious of these is that
the equation (and therefore its solution) has an irreparable divergence,
irreparable in the sense that it persists in a cut-off model with finite
bath relaxation time 
\cite{ford01d}. We believe this divergence is related to a well known
phenomenon of quantum field theory: the Hilbert space spanned by the
eigenfunctions of the coupled Hamiltonian is orthogonal to that of the
uncoupled Hamiltonian \cite{vanhove52}. It can be shown that this is the
case for the system of coupled oscillators that is the microscopic basis
for the derivation of the exact master equation and its solution \cite
{araki82,efimov94}.

The exception is the high temperature limit, where the divergence does not
appear since, by convention, one there neglects the zero-point
oscillations of the bath. We therefore confine our discussion to that
limit. However, even in that case there are further difficulties. The
first of these is that the initial state breaks the translational
invariance of the system. This is manifest in an initial squeeze centered
at the origin that, for a free particle, results in a drift of the mean
position toward the origin. For this reason in the examples we consider
only the case of a free particle, where this difficulty is most apparent.

A second difficulty with the equation in the high temperature limit is
that for short times the diagonal elements of the density matrix, which
have the physical interpretation of probabilities, are not necessarily
positive. This is a well known difficulty with the weak coupling master
equation, solved by a further time average to bring it to the so-called
Lindblad form that guarantees positivity \cite{lindblad}. It was supposed
that the exact master equation, with its time-dependent coefficients,
would be immune but, as we shall show explicitly, in the high temperature
limit this difficulty remains. The situation, therefore, is that there
are concerns that seriously limit the utility of the exact master
equation. The irreparable divergence can only be avoided with the neglect
of zero-point oscillations in the high temperature limit, and even there
problems remain with the failure of translational invariance and
positivity.

Despite these problems, we argue that the solution of the exact master
equation can be useful. In particular, we show that when the initial
particle temperature is taken to be equal to that of the bath, we obtain
results that for short times are identical with those obtained \ from a
calculation which considers entanglement at all times \cite{ford01a}.
Finally, an important result is the demonstration that measures of
decoherence based on the Wigner function are identical with those based on
the off-diagonal elements of the density matrix. We argue that the Wigner
function, being everywhere real, is the preferable description.

\section{General solution in high temperature limit}

The exact master equation, in terms of the Wigner function $W(q,p,t)$ for
the complete system of oscillator plus heat bath, may be expressed in the
form

\begin{eqnarray}
\frac{\partial W}{\partial t} &=& -\frac{1}{m}p\frac{\partial W}{\partial
q}+m\Omega^{2}(t)q\frac{\partial W}{\partial p} \nonumber \\
&&{}+2\Gamma(t)\frac{\partial pW}{\partial p}+\hbar m\Gamma
(t)h(t)\frac{\partial^{2}W}{\partial p^{2}}+\hbar\Gamma
(t)f(t)\frac{\partial^{2} W}{\partial q\partial p},
\label{2.1}
\end{eqnarray}
and we note the presence of four time-dependent coefficients for which we
obtained explicit expressions \cite{ford01d}.
 
The general solution of the exact master
equation is most simply expressed in terms of the Wigner characteristic
function (the Fourier transform of the Wigner function).
\begin{equation}
\tilde{W}(Q,P;t)=\int dq\int dpe^{-i(qP+pQ)/\hbar }W(q,p;t).  \label{2.2}
\end{equation} Expressed in terms of this function, the general solution
given in Eq. (4.15) of \cite{ford01d} takes the simple form,
\begin{eqnarray}
\tilde{W}(Q,P;t) &=&\exp \{-\frac{\left\langle X^{2}\right\rangle
P^{2}+m\left\langle X\dot{X}+\dot{X}X\right\rangle QP+m^{2}\left\langle
\dot{ X}^{2}\right\rangle Q^{2}}{2\hbar ^{2}}\}  \nonumber \\ 
&&\times
\tilde{W}(m\dot{G}Q+GP,m^{2}\ddot{G}Q+m\dot{G}P;0).  \label{2.3}
\end{eqnarray} In this expression $X(t)$ is the fluctuating position
operator, defined in Eq. (2.16) of \cite{ford01d}, while $G(t)$ is the
Green function, defined in general to be
\begin{equation} G(t_{2}-t_{1})=\frac{1}{i\hbar
}[x(t_{1}),x(t_{2})]\theta (t_{2}-t_{1}),
\label{2.4}
\end{equation} in which $x(t)$ is the time-dependent Heisenberg
coordinate operator and $
\theta $ is the Heaviside function.

It was shown in \cite{ford01d} (Sec. V.A.2) that the quadratic
expectations
$
\left\langle X^{2}\right\rangle $, $\left\langle X\dot{X}+\dot{X}
X\right\rangle $ and $\left\langle \dot{X}^{2}\right\rangle $ are
logarithmically divergent, which has the effect that the solution
(\ref{2.3}) vanishes identically for non-zero positive times. There,
too, it was shown that the divergence is irreparable, in the sense that
it persists in a model with a high frequency cut-off. As was stated in
the Introduction, it appears that this is a reflection of the well
established fact that the states of the uncoupled system (the initial
state) are orthogonal to those of the coupled system (states for later
times) \cite{araki82,efimov94}. This fact, surprising from the point of
view of the quantum mechanics of finite systems, is a feature only of
systems with an infinite number of degrees of freedom. In our case it is
the heat bath that must be infinite. In any event, the upshot is that,
while the derivation of the exact master equation is formally correct,
there is no useful result. The exception is the high temperature limit,
where by convention one neglects zero-point oscillations. We therefore
restrict our further discussion to this limit. Indeed, it is this limit
(or better said: approximation) that has been considered in all previous
discussions of the exact master equation.

In considering the result in the high temperature limit, we restrict our
discussion to the Ohmic model of coupling to the heat bath, which is
adequate for our purposes. This model corresponds to Newtonian friction,
with retarding force proportional to the velocity. In addition we consider
only the case of a free particle, moving in the absence of any external
potential. Choosing the friction constant to be $m\gamma $, the Green
function then takes the simple form:
\begin{equation} G(t)=\frac{1-e^{-\gamma t}}{m\gamma }.  \label{2.5}
\end{equation} In the high temperature limit the moments appearing in the
solution can then be written in the form:
\begin{eqnarray}
\left\langle X^{2}\right\rangle &=&2m\gamma kT\int_{0}^{t}dt^{\prime
}G^{2}(t^{\prime })  \nonumber \\ &=&\frac{kT}{m\gamma ^{2}}[2\gamma
t-(1-e^{-\gamma t})(3-e^{-\gamma t})], 
\nonumber \\
\left\langle X\dot{X}+\dot{X}X\right\rangle &=&2m\gamma kTG^{2}(t) 
\nonumber \\ &=&\frac{2kT}{m\gamma }(1-e^{-\gamma t})^{2},  \nonumber \\
\left\langle \dot{X}^{2}\right\rangle &=&2m\gamma kT\int_{0}^{t}dt^{\prime
}
\dot{G}^{2}(t^{\prime })  \nonumber \\ &=&\frac{kT}{m}(1-e^{-2\gamma t}). 
\label{2.6}
\end{eqnarray} We should emphasize that in these \ expressions $T$ is the
temperature of the bath. In the initial state the bath and the particle
are uncoupled, so in the general solution (\ref{2.3}) $\tilde{W}(Q,P;0)$,
the initial Wigner characteristic function, may be chosen to have any
form consistent with a single particle not coupled to a bath. In the
examples we consider two possibilities: the particle initially at zero
temperature and at a temperature equal to that of the bath.

\section{Example: Gaussian wave packet}

\subsection{Initial particle temperature zero}

We consider first an initial state corresponding to a Gaussian wave
packet, with wave function of the form
\begin{equation}
\phi (x,0)=\frac{1}{(2\pi \sigma ^{2})^{1/4}}\exp \{-\frac{(x-x_{0})^{2}}{
4\sigma ^{2}}\}.  \label{3.1}
\end{equation} This is a stationary wave packet, centered at $x_{0}$ with
mean square width 
$\sigma ^{2}$. It is a minimum uncertainty wave packet, since the mean
square momentum has the minimum value $\hbar ^{2}/4\sigma ^{2}$. The
corresponding Wigner characteristic function is
\begin{eqnarray}
\tilde{W}(Q,P;0) &=&\int_{-\infty }^{\infty }dxe^{-iPx/\hbar }\phi
(x-\frac{Q }{2},0)\phi ^{\ast }(x+\frac{Q}{2},0)  \nonumber \\ &=&\exp
\{-\frac{Q^{2}}{8\sigma ^{2}}-\frac{\sigma ^{2}P^{2}}{2\hbar ^{2}}-i
\frac{x_{0}P}{\hbar }\}.  \label{3.2}
\end{eqnarray}

This initial state is what is termed a pure state. The condition for a
pure state is usually stated in terms of the density matrix:
$\mathrm{Tr}\{\rho ^{2}\}=1$. In terms of the Wigner characteristic
function, this condition becomes
\begin{equation}
\frac{1}{2\pi \hbar }\int_{-\infty }^{\infty }dQ\int_{-\infty }^{\infty
}dP\left\vert \tilde{W}(Q,P)\right\vert ^{2}=1.  \label{3.3}
\end{equation} It is easy to show that a state is a pure state if and
only if it can be associated to a wave function. Such a pure state
necessarily corresponds to a particle at zero temperature; a
finite-temperature state is a mixed state.

Putting the expression (\ref{3.2}) for the initial Wigner characteristic
function in the general solution (\ref{2.3}) we see that
\begin{equation}
\tilde{W}(Q,P;t)=\exp \{-\frac{
A_{11}^{(0)}P^{2}+2A_{12}^{(0)}PQ+A_{22}^{(0)}Q^{2}}{2\hbar ^{2}}-i\frac{
x_{0}(m^{2}\ddot{G}Q+m\dot{G}P)}{\hbar }\},  \label{3.4}
\end{equation} where we have introduced
\begin{eqnarray} A_{11}^{(0)} &=&\left\langle X^{2}\right\rangle +\sigma
^{2}m^{2}\dot{G}^{2}+
\frac{\hbar ^{2}G^{2}}{4\sigma ^{2}}  \nonumber \\ &=&\frac{kT}{m\gamma
^{2}}[2\gamma t-(1-e^{-\gamma t})(3-e^{-\gamma t})]+\sigma
^{2}e^{-2\gamma t}+\frac{\hbar ^{2}(1-e^{-\gamma t})^{2}}{ 4m^{2}\gamma
^{2}\sigma ^{2}},  \nonumber \\ A_{12}^{(0)} &=&m\frac{\left\langle
X\dot{X}+\dot{X}X\right\rangle }{2} +\sigma
^{2}m^{3}\dot{G}\ddot{G}+\frac{\hbar ^{2}mG\dot{G}}{4\sigma ^{2}} 
\nonumber \\ &=&\frac{kT}{\gamma }(1-e^{-\gamma t})^{2}-m\gamma \sigma
^{2}e^{-2\gamma t}+
\frac{\hbar ^{2}(e^{-\gamma t}-e^{-2\gamma t})}{4m\gamma \sigma ^{2}}, 
\nonumber \\ A_{22}^{(0)} &=&m^{2}\left\langle \dot{X}^{2}\right\rangle
+\sigma ^{2}m^{4}
\ddot{G}^{2}+\frac{\hbar ^{2}m^{2}\dot{G}^{2}}{4\sigma ^{2}}  \nonumber \\
&=&mkT(1-e^{-2\gamma t})+m^{2}\gamma ^{2}\sigma ^{2}e^{-2\gamma t}+\frac{
\hbar ^{2}e^{-2\gamma t}}{4\sigma ^{2}}.  \label{3.5}
\end{eqnarray} Here we have used the superscript \textquotedblleft
$(0)$\textquotedblright\ to indicate that the particle is initially at
zero temperature before suddenly being coupled to the bath at temperature
$T$.

We now form the Wigner function with the inverse Fourier transform
\begin{equation} W(q,p;t)=\frac{1}{(2\pi \hbar )^{2}}\int dQ\int
dPe^{i(qP+pQ)/\hbar }\tilde{W }(Q,P;t).  \label{3.6}
\end{equation} With the expression (\ref{3.4}) for $\tilde{W}(Q,P;t)$,
the integration is a standard Gaussian integral \cite{ford01d}, with the
result
\begin{equation}
W(q,p;t)=W^{(0)}(q-m\dot{G}x_{0},p-m^{2}\ddot{G}x_{0};t),  \label{3.7}
\end{equation} where
\begin{equation} W^{(0)}(q,p;t)=\frac{1}{2\pi
\sqrt{A_{11}^{(0)}A_{22}^{(0)}-A_{12}^{(0)2}}}
\exp \{-\frac{A_{22}^{(0)}q^{2}-2A_{12}^{(0)}qp+A_{11}^{(0)}p^{2}}{
2(A_{11}^{(0)}A_{22}^{(0)}-A_{12}^{(0)2})}\}  \label{3.8}
\end{equation} is the Wigner function corresponding to a single minimum
uncertainty wave packet initially at the origin.

The Wigner distribution (\ref{3.7}) corresponds to a Gaussian distribution
in phase space, centered at the point $(\left\langle x(t)\right\rangle
,\left\langle p(t)\right\rangle )$, where
\begin{eqnarray}
\left\langle x(t)\right\rangle &=&m\dot{G}x_{0}=x_{0}e^{-\gamma t}, 
\nonumber
\\
\left\langle p(t)\right\rangle &=&m^{2}\ddot{G}x_{0}=-m\gamma
x_{0}e^{-\gamma t}.  \label{3.9}
\end{eqnarray} The width of the distribution is characterized by the
coefficients $ A_{jk}^{(0)}$, which have the interpretation
\begin{eqnarray} A_{11}^{(0)} &=&\Delta x^{2}\equiv \left\langle
x^{2}(t)\right\rangle -\left\langle x(t)\right\rangle ^{2},  \nonumber \\
A_{22}^{(0)} &=&\Delta p^{2}\equiv \left\langle p^{2}(t)\right\rangle
-\left\langle p(t)\right\rangle ^{2},  \nonumber \\ A_{12}^{(0)}
&=&\frac{1}{2}\left\langle x(t)p(t)+p(t)x(t)\right\rangle -\left\langle
x(t)\right\rangle \left\langle p(t)\right\rangle .
\label{3.10}
\end{eqnarray} The breaking of translational invariance is evident, since
the center of the distribution drifts to the origin, which is a special
point. We get more insight into this phenomenon by forming the limit as
$t$ goes to zero through positive values,
\begin{equation} W(q,p;0^{+})=\frac{1}{\pi \hbar }\exp
\{-\frac{(q-x_{0})^{2}}{2\sigma ^{2}}-
\frac{2\sigma ^{2}(p+m\gamma q)^{2}}{\hbar ^{2}}\}.  \label{3.11}
\end{equation} Here we see that the particle has received an impulse
$-m\gamma q$ proportional to the displacement and directed toward the
origin. This is exactly the action that produces a squeezed state
\cite{garrett97,ford02}. Indeed, as one can readily verify, the Wigner
function (\ref{3.11}) corresponds to a pure state with associated wave
function
\begin{equation}
\phi (x,0^{+})=\frac{1}{(2\pi \sigma ^{2})^{1/4}}\exp
\{-\frac{(x-x_{0})^{2} }{4\sigma ^{2}}-i\frac{m\gamma }{2\hbar }x^{2}\}. 
\label{3.12}
\end{equation} Thus, the state immediately after the coupling to the bath
is still a zero temperature pure state. However, as a result of the
squeeze toward the origin, it is no longer a minimum uncertainty state.
It still corresponds to a wave packet centered at $x_{0}$ with mean
square width $\sigma ^{2}$, but the mean square momentum uncertainty is
now $\Delta p^{2}=\hbar ^{2}/4\sigma ^{2}+m^{2}\gamma ^{2}\sigma ^{2}$,
greater than the minimum value. The result of the squeeze is that the
distribution (\ref{3.7}) drifts toward the origin. Less obvious perhaps,
the same phenomenon appears in a shrinking width of the probability
distribution at short times. At longer times the width expands due to
thermal spreading

\subsection{Initial particle temperature equal to that of the bath}

As we have noted, the minimum uncertainty wave packet (\ref{3.1})
considered above corresponds to a particle at temperature zero. A perhaps
more realistic initial state would be that for a particle initially at
temperature $T$, equal to that of the bath. In \cite{ford01d} we showed
that the initial state at temperature $T$ is obtained from that at
temperature zero by the replacement
\begin{equation}
\tilde{W}(Q,P;0)\rightarrow \exp \{-\frac{mkTQ^{2}}{2\hbar
^{2}}\}\tilde{W} (Q,P;0).  \label{3.13}
\end{equation} This finite temperature state is a mixed state, no longer
satisfying the condition (\ref{3.3}) for a pure state. It cannot be
associated to a wave function. Making this replacement with the zero
temperature form (\ref{3.2}) and repeating the steps leading to the
result (\ref{3.7}), we obtain
\begin{equation}
W(q,p;t)=W^{(T)}(q-m\dot{G}x_{0},p-m^{2}\ddot{G}x_{0};t),  \label{3.14}
\end{equation} where now the Wigner function corresponding to a single
Gaussian wave packet initially at the origin takes the form
\begin{equation} W^{(T)}(q,p;t)=\frac{1}{2\pi
\sqrt{A_{11}^{(T)}A_{22}^{(T)}-A_{12}^{(T)2}}}
\exp \{-\frac{A_{22}^{(T)}q^{2}-2A_{12}^{(T)}qp+A_{11}^{(T)}p^{2}}{
2(A_{11}^{(T)}A_{22}^{(T)}-A_{12}^{(T)2})}\},  \label{3.15}
\end{equation} in which
\begin{eqnarray} A_{11}^{(T)} &=&A_{11}^{(0)}+mkTG^{2}  \nonumber \\
&=&\frac{2kT}{m\gamma ^{2}}(\gamma t-1+e^{-\gamma t})+\sigma
^{2}e^{-2\gamma t}+\frac{\hbar ^{2}(1-e^{-\gamma t})^{2}}{4m^{2}\gamma
^{2}\sigma ^{2}}, 
\nonumber \\ A_{12}^{(T)} &=&A_{12}^{(0)}+m^{2}kTG\dot{G}  \nonumber \\
&=&\frac{kT}{\gamma }(1-e^{-\gamma t})-m\gamma \sigma ^{2}e^{-2\gamma t}+
\frac{\hbar ^{2}e^{-\gamma t}(1-e^{-\gamma t})}{4m\gamma \sigma ^{2}}, 
\nonumber \\ A_{22}^{(T)} &=&A_{22}^{(0)}+m^{3}kT\dot{G}^{2}  \nonumber \\
&=&mkT+m^{2}\gamma ^{2}\sigma ^{2}e^{-2\gamma t}+\frac{\hbar
^{2}e^{-2\gamma t}}{4\sigma ^{2}}.  \label{3.16}
\end{eqnarray} Note that the motion of the center of the distribution is
the same as that given in Eq. (\ref{3.9}) for the case of initial
particle temperature zero. The only change is an increase in the width of
the Gaussian distribution due to thermal spreading, the widths being
given by the expressions (\ref{3.10}) with $A_{jk}^{(T)}$ in place of
$A_{jk}^{(0)}$. We again get more insight into this by forming the limit
as $t$ goes to zero through positive values,
\begin{equation} W(q,p;0^{+})=\frac{1}{\pi \hbar \sqrt{1+4\frac{\sigma
^{2}}{\lambda _{T}^{2}} }}\exp \{-\frac{(q-x_{0})^{2}}{2\sigma
^{2}}-\frac{(p+m\gamma q)^{2}}{2(
\frac{\hbar ^{2}}{4\sigma ^{2}}+mkT)}\}.  \label{3.17}
\end{equation} Here we see that that there is the same initial squeeze as
in Eq. (\ref{3.11} ), but the squeeze acts on a state with mean square
momentum uncertainty increased by the mean square thermal momentum,
$mkT$. The result is that immediately after the squeeze the mean square
momentum uncertainty is $
\Delta p^{2}=\hbar ^{2}/4\sigma ^{2}+mkT+m^{2}\gamma ^{2}\sigma ^{2}$.
However, the initial $\Delta x^{2}=\sigma ^{2}$ and is unaffected.by the
squeeze.

\section{Example: Pair of Gaussian wave packets}

\subsection{Initial particle temperature zero}

The wave function for an initial "Schr\"{o}dinger cat" state consisting of
two separated minimum uncertainty Gaussian wave packets has the form
\begin{equation}
\phi (x,0)=\frac{1}{(8\pi \sigma ^{2})^{1/4}(1+e^{-d^{2}/8\sigma
^{2}})^{1/2} }(\exp \{-\frac{(x-\frac{d}{2})^{2}}{4\sigma ^{2}}\}+\exp
\{-\frac{(x+\frac{d }{2})^{2}}{4\sigma ^{2}}\}).  \label{4.1}
\end{equation} The initial Wigner characteristic function is then given by
\begin{equation}
\tilde{W}(Q,P;0)=\frac{1}{1+e^{-d^{2}/8\sigma ^{2}}}\exp \{-\frac{Q^{2}}{
8\sigma ^{2}}-\frac{\sigma ^{2}P^{2}}{2\hbar ^{2}}\}(\cos \frac{Pd}{2\hbar
} +e^{-d^{2}/8\sigma ^{2}}\cosh \frac{Qd}{4\sigma ^{2}}).  \label{4.2}
\end{equation} Therefore, from the general solution (\ref{2.3}) we can
write,
\begin{eqnarray}
\tilde{W}(Q,P;t) &=&\frac{1}{1+e^{-d^{2}/8\sigma ^{2}}}\exp \{-\frac{
A_{11}^{(0)}P^{2}+2A_{12}^{(0)}PQ+A_{22}^{(0)}Q^{2}}{2\hbar ^{2}}\} 
\nonumber
\\ &&\times (\cos \frac{(m^{2}\ddot{G}Q+m\dot{G}P)d}{2\hbar
}+e^{-d^{2}/8\sigma ^{2}}\cosh \frac{(m\dot{G}Q+GP)d}{4\sigma ^{2}}). 
\label{4.3}
\end{eqnarray} Here the $A_{ij}^{(0)}$ are again given by (\ref{3.5}).

The inverse Fourier transform (\ref{3.6}) again involves only standard
Gaussian integrals. We thus easily see that the Wigner function for a
wave-packet pair is
\begin{eqnarray} W(q,p;t) &=&\frac{1}{2(1+e^{-d^{2}/8\sigma ^{2}})}{\LARGE
\{}W^{(0)}(q-\frac{ m\dot{G}d}{2},p-\frac{m^{2}\ddot{G}d}{2})  \nonumber \\
&&+W^{(0)}(q+\frac{m\dot{G}d}{2},p+\frac{m^{2}\ddot{G}d}{2})  \nonumber \\
&&+2e^{-A^{(0)}(t)}W^{(0)}(q,p)\cos \Phi ^{(0)}(q,p:t){\LARGE \}},
\label{4.4}
\end{eqnarray} where $W^{(0)}$ is the Wigner function (\ref{3.8}) for a
single minimum uncertainty wave packet at the origin and we have
introduced
\begin{eqnarray} A^{(0)}(t) &=&\frac{(A_{11}^{(0)}-\frac{\hbar
^{2}G^{2}}{4\sigma ^{2}} )(A_{22}^{(0)}-\frac{\hbar
^{2}m^{2}\dot{G}^{2}}{4\sigma ^{2}} )-(A_{12}^{(0)}-\frac{\hbar
^{2}mG\dot{G}}{4\sigma ^{2}})^{2}}{
A_{11}^{(0)}A_{22}^{(0)}-A_{12}^{(0)2}}\frac{d^{2}}{8\sigma ^{2}}  \nonumber
\\
\Phi ^{(0)}(q,p;t)
&=&\frac{(GA_{22}^{(0)}-m\dot{G}A_{12}^{(0)})q+(m\dot{G}
A_{11}^{(0)}-GA_{12}^{(0)})p}{A_{11}^{(0)}A_{22}^{(0)}-A_{12}^{(0)2}}\frac{
\hbar d}{4\sigma ^{2}}.  \label{4.5}
\end{eqnarray} The first two terms in brackets in (\ref{4.4}) correspond
to a pair of Gaussian wave packets of the form (\ref{3.7}), centered
initially at $ x_{0}=\pm d/2$ and propagating independently. We can call
these the direct terms. The third term is an interference term. The
direct terms are a maximum at their centers, which are drifting toward
the origin but for $d\gg
\sigma $ will for short times be well away from the origin. The
interference term is a maximum at the origin and initially its peak
height is exactly twice that of either of the direct terms. (However, as
one can readily verify, the area under the interference term is
independent of time and a factor $e^{-d^{2}/8\sigma ^{2}}$ smaller than
the area under the direct terms.) The factor $e^{-A^{(0)}(t)}$, equal to
the\ ratio of the peak height of the interference term to twice the peak
height of the either of the direct terms, is taken as a measure of the
interference in phase space. In Fig. 1 we show $A^{(0)}$ as a function of
$\gamma t$ and there we see that for $\gamma t\ll 1$, $A^{(0)}$ is linear
in $t$. Using the expressions (\ref {3.5}) for the quantities
$A_{11}^{(0)}$ and (\ref{2.5}) for $G$ we can readily show that for short
times
\begin{equation} A^{(0)}(t)\cong \frac{d^{2}}{\lambda
_{\mathrm{th}}^{2}}\gamma t,
\label{4.6}
\end{equation} where $\lambda _{\mathrm{th}}$ is the mean de Broglie
wavelength,
\begin{equation}
\lambda _{\mathrm{th}}=\frac{\hbar }{\sqrt{mkT}}.  \label{4.7}
\end{equation} For a separation large compared with the thermal de
Broglie wave length, this corresponds to a decay of the peak height of
the interference term on a time scale short compared with $\gamma ^{-1}$,
\begin{equation}
\tau _{d}=\frac{\lambda _{\mathrm{th}}^{2}}{d^{2}}\gamma ^{-1}. 
\label{4.8}
\end{equation} This is the frequently appearing "decoherence time"
\cite{ford01a}.

It is of interest to form the probability distribution, which in general
is given by
\begin{eqnarray} P(x;t) &=&\int_{-\infty }^{\infty }dpW(q,p;t)  \nonumber \\
&=&\frac{1}{2\pi \hbar }\int_{-\infty }^{\infty }dPe^{ixP/\hbar }\tilde{W}
(0,P;t).  \label{4.9}
\end{eqnarray} Using the expression (\ref{4.3}) for the Wigner
characteristic function, this can be written
\begin{eqnarray} P(x;t) &=&\frac{1}{2(1+e^{-d^{2}/8\sigma
^{2}})}[P^{(0)}(x-m\dot{G}\frac{d}{2 })+P^{(0)}(x+m\dot{G}\frac{d}{2}) 
\nonumber \\ &&+2a(t)\exp
\{-\frac{m^{2}\dot{G}^{2}d^{2}}{8A_{11}^{(0)}}\}P^{(0)}(x)\cos 
\frac{\hbar Gdx}{4\sigma ^{2}A_{11}^{(0)}}),  \label{4.10}
\end{eqnarray} where
\begin{equation} P^{(0)}(x)=\frac{1}{\sqrt{2\pi A_{11}^{(0)}}}\exp
\{-\frac{x^{2}}{ 2A_{11}^{(0)}}\}  \label{4.11}
\end{equation} is the probability distribution for a single minimum
uncertainty wave packet initially centered a the origin and we have
introduced the attenuation coefficient,
\begin{equation} a(t)=\exp \{-\frac{\left\langle X^{2}\right\rangle
d^{2}}{8\sigma ^{2}A_{11}^{(0)}}\}.  \label{4.12}
\end{equation} The attenuation coefficient is a measure of the
interference as it would be observed in the probability distribution. We
should emphasize that the probability distribution can be directly
measured. This is to be contrasted with the Wigner distribution in phase
space, which is not directly observable.

The integrated probability in the interference term is equal to $
e^{-d^{2}/8\sigma ^{2}}$, independent of time. For $d\gg \sigma $ this is
negligibly small. But it is the \emph{amplitude} of the interference
fringes that is observed and measured by the attenuation coefficient
(\ref{4.12}).

For short times, $\gamma t\ll 1$, using the expressions (\ref{2.6}) for $
\left\langle X^{2}\right\rangle $ and (\ref{3.5}) for $A_{11}^{(0)}$, we
see that the attenuation coefficient takes the form
\begin{equation} a(t)\cong \exp \{-\frac{t^{3}}{3\tau
_{d}[t^{2}+(2m\sigma ^{2}/\hbar )^{2}]}
\}.  \label{4.13}
\end{equation} This initially falls very slowly from unity, but after a
time $2m\sigma ^{2}/\hbar $ (the time for the mean square width $\Delta
x^{2}$ to double) will decay with a characteristic time $3\tau _{d}$,
where $\tau _{d}$ is the characteristic time (\ref{4.8}) for the decay of
the interference term in the Wigner function.

With no coupling to the bath the interference pattern in the probability
distribution would not appear until the direct terms begin to overlap due
to wave packet spreading. Taking $\gamma \rightarrow 0$ in the expression
(\ref {3.5}) for the mean square width, we see that this would be a time
of order $ t_{\mathrm{mix}}=2m\sigma d/\hbar \ll \tau _{d}$. What this
means is that neither the rapid disappearance of the interference term in
the Wigner distribution, nor the corresponding rapid decay of the
attenuation coefficient, can could be directly observed. All that can be
seen is the non-appearance of the interference term in the probability
distribution.

\subsection{Initial particle temperature equal to that of the bath}

The corresponding results for the particle initially at the temperature
$T$ of the bath are obtained using the prescription (\ref{3.13}). The
result is
\begin{eqnarray} W(q,p;t) &=&\frac{1}{2(1+e^{-d^{2}/8\sigma ^{2}})}\left\{
W^{(T)}(q-\frac{m
\dot{G}d}{2},p-\frac{m^{2}\ddot{G}d}{2})\right.  \nonumber \\
&&+W^{(T)}(q+\frac{m\dot{G}d}{2},p+\frac{m^{2}\ddot{G}d}{2})  \nonumber \\
&&\left. +2e^{-A^{(T)}(t)}W^{(T)}(q,p)\cos \Phi ^{(T)}(q,p:t)\right\} ,
\label{4.14}
\end{eqnarray} where
\begin{eqnarray} A^{(T)}(t) &=&\frac{(A_{11}^{(T)}-\frac{\hbar
^{2}G^{2}}{4\sigma ^{2}} )(A_{22}^{(T)}-\frac{\hbar
^{2}m^{2}\dot{G}^{2}}{4\sigma ^{2}} )-(A_{12}^{(T)}-\frac{\hbar
^{2}mG\dot{G}}{4\sigma ^{2}})^{2}}{
A_{11}^{(T)}A_{22}^{(T)}-A_{12}^{(T)2}}\frac{d^{2}}{8\sigma ^{2}}  \nonumber
\\
\Phi ^{(T)}(q,p;t)
&=&\frac{(GA_{22}^{(T)}-m\dot{G}A_{12}^{(T)})q+(m\dot{G}
A_{11}^{(T)}-GA_{12}^{(T)})p}{A_{11}^{(T)}A_{22}^{(T)}-A_{12}^{(T)2}}\frac{
\hbar d}{4\sigma ^{2}}.  \label{4.15}
\end{eqnarray} This expression for the Wigner function has the same form
as that found above for the case where the initial particle temperature
is zero. The only difference is that $A_{jk}^{(0)}\rightarrow
A_{jk}^{(T)}$. The area under the interference term is still independent
of time and smaller than the area under the direct terms by the same
factor $e^{-d^{2}/8\sigma ^{2}}$. The main difference is that the peak
height of the interference term is much reduced. The ratio of the peak
height of the interference term to twice that of the direct terms is
given by the factor $e^{-A^{(T)}(t)}$ but, as we see in Fig. 1, $A^{(T)}$
no longer vanishes at $t=0$. Instead, we now find for
$
\gamma t\ll 1$,
\begin{equation} A^{(T)}(t)\cong \frac{d^{2}}{2\lambda
_{\mathrm{th}}^{2}+8\sigma ^{2}}.
\label{4.16}
\end{equation} For separation large compared with both the thermal de
Broglie wavelength and the slit width, this will be a large number and
the peak height of the interference term will be correspondingly small.
Thus, we see that there is no time scale for the disappearance of the
interference term, it is already negligibly small at $t=0$. We might say
that this is just the point. What has occurred is that an incoherent
superposition of pure states has wiped out the interference term. In the
present case this is the result of the initial preparation of the state.
In the previous case of an initial pure state, this occurs as a result of
the "warming up" of the particle in a time 
$\tau _{d}$. In either case the interference term is gone long before its
effect could be observed in the probability distribution.

The probability distribution is of the same form (\ref{4.10}), but with $
A_{11}^{(T)}(t)$ replacing $A_{11}^{(0)}(t)$. The attenuation coefficient
is now given by
\begin{equation} a(t)=\exp \{-\frac{(\left\langle X^{2}\right\rangle
+mkTG^{2})d^{2}}{8\sigma ^{2}A_{11}^{(T)}}\}  \label{4.17}
\end{equation} For $\gamma t\ll 1$ this becomes
\begin{equation} a(t)\cong \exp \{-\frac{kTd^{2}}{8m\sigma ^{4}}t^{2}\} 
\label{4.18}
\end{equation} These results for short times are identical with those
obtained from an exact calculation assuming entanglement of the particle
with the bath at all times \cite{ford01a}). Since the coupling to the
bath, as measured by the decay rate $\gamma $, does not appear they can
also be obtained from a calculation without dissipation using only the
familiar formulas of elementary quantum mechanics \cite{murakami03}. Thus
we see that, while the exact master equation gives here correct results,
it does so only for times so short that coupling to the bath can be
neglected.

\section{Density matrix}

The density matrix is a Hermitian operator, $\rho $, defined such that for
any wave function $\psi $ corresponding to a state of the particle in the
Hilbert space $\left\langle \psi ,\rho \psi \right\rangle $ is the
probability that the system is in the state. From the notion of
probability, we see immediately the the density matrix must be a positive
definite operator. Consider the eigenfunctions $\phi _{a}$ and the
corresponding eigenvalues $p_{a}$ of the density matrix, 
\begin{equation}
\rho \phi _{a}=p_{a}\phi _{a}.  \label{5.1}
\end{equation} Clearly, the eigenvalue $p_{a}$ is the probability that
the system is in state $\phi _{a}$ and is therefore necessarily positive.
We assume the density matrix is normalized, so that
\begin{equation}
\sum_{a}p_{a}=1.  \label{5.2}
\end{equation} This is a sum of positive terms equal to unity. It follows
immediately that
\begin{equation}
\mathrm{Tr}\{\rho ^{2}\}=\sum_{a}p_{a}^{2}\leq 1.  \label{5.3}
\end{equation} Moreover, the equality holds if and only if there is
exactly one eigenstate with probability one, all others having
probability zero. This is exactly the condition (\ref{3.3}) for a pure
state.

In coordinate space the elements of the density matrix are given in terms
of the Wigner characteristic function by the formula:
\begin{eqnarray}
\left\langle x\left\vert \rho \right\vert x^{\prime }\right\rangle
&=&\int_{-\infty }^{\infty }dpe^{i(x-x^{\prime })p/\hbar }W(\frac{
x+x^{\prime }}{2},p)  \nonumber \\ &&\frac{1}{2\pi \hbar }\int_{-\infty
}^{\infty }dPe^{i(x+x^{\prime })P/2\hbar }\tilde{W}(x^{\prime }-x,P). 
\label{5.4}
\end{eqnarray} The normalization condition (\ref{5.2}) becomes
\begin{equation}
\mathrm{Tr}\{\rho \}=\int_{-\infty }^{\infty }dx\left\langle x\left\vert
\rho \right\vert x\right\rangle =\tilde{W}(0,0)=1.  \label{5.5}
\end{equation} That is, the Wigner characteristic function corresponding
to a normalized density matrix must have the value unity at the origin.

Next consider
\begin{eqnarray}
\mathrm{Tr}\{\rho ^{2}\} &=&\int_{-\infty }^{\infty }dx\int_{-\infty
}^{\infty }dx^{\prime }\left\langle x\left\vert \rho \right\vert x^{\prime
}\right\rangle \left\langle x^{\prime }\left\vert \rho \right\vert
x\right\rangle  \nonumber \\ &=&\frac{1}{2\pi \hbar }\int_{-\infty }^{\infty
}dQ\int_{-\infty }^{\infty }dP\left\vert \tilde{W}(Q,P)\right\vert ^{2}, 
\label{5.7}
\end{eqnarray} where we have used that fact that
$\tilde{W}(-Q,-P)=\tilde{W}(Q,P)^{\ast }$. Thus, the condition
(\ref{5.3}) takes the form
\begin{equation}
\frac{1}{2\pi \hbar }\int_{-\infty }^{\infty }dQ\int_{-\infty }^{\infty
}dP\left\vert \tilde{W}(Q,P)\right\vert ^{2}\leq 1,  \label{5.8}
\end{equation} with the equality holding if and only if the density
matrix is that of a pure state. This is the condition (\ref{3.3}) for a
pure state.

\subsection{Gaussian wave packet}

Consider now the case of an initial state corresponding to a single
minimum uncertainty wave packet, for which the Wigner characteristic
function at time $t$ is given in Eq. (\ref{3.4}). The evaluation of the
formula (\ref {5.4}) for the matrix element of the density operator
involves a single Gaussian integral. The result can be written
\begin{equation}
\left\langle x\left\vert \rho (t)\right\vert x^{\prime }\right\rangle
=\exp
\{i\frac{m^{2}\ddot{G}}{\hbar }x_{0}(x-x^{\prime })\}R(x-m\dot{G}
x_{0},x^{\prime }-m\dot{G}x_{0};t),  \label{5.8a}
\end{equation} where we have introduced
\begin{equation} R(x,x^{\prime };t)=\frac{1}{\sqrt{2\pi A_{11}^{(0)}}}\exp
\{-\frac{\frac{4}{
\hbar ^{2}}(A_{11}^{(0)}A_{22}^{(0)}-A_{12}^{(0)2})(x-x^{\prime
})^{2}+(x+x^{\prime })^{2}}{8A_{11}^{(0)}}+i\hbar \frac{
A_{12}^{(0)}(x^{2}-x^{\prime 2})}{2\hbar ^{2}A_{11}^{(0)}}\}, 
\label{5.8b}
\end{equation} which is the matrix element corresponding to a single wave
packet initially centered at the origin. Using the expressions
(\ref{2.5}) for the Green function and (\ref{3.5}) for $A_{jk}^{(0)}(t)$,
we see that viewed in the
$ xx^{\prime }$ plane the absolute value of the density matrix,
$\left\vert
\left\langle x\left\vert \rho (t)\right\vert x^{\prime }\right\rangle
\right\vert $, consists of a single Gaussian peak. This peak is initially
centered at $x=x^{\prime }=x_{0}$ and circularly symmetric with width $
\sigma $. In the course of time the center of the peak drifts to the
origin while stretching in the direction $x=x^{\prime }$.

For short times, the density matrix is not necessarily positive. The
simplest way to see this is to form $\mathrm{Tr}\{\rho ^{2}\}$, using the
above expression for the density matrix element or evaluating directly the
expression (\ref{5.7}). In either case we find
\begin{equation}
\mathrm{Tr}\{\rho ^{2}\}=\frac{\hbar }{2\sqrt{
A_{11}^{(0)}A_{22}^{(0)}-A_{12}^{(0)2}}}.  \label{5.9}
\end{equation} In Fig. 2 we show a plot over a short initial time
interval of this result, evaluated using the expressions (\ref{3.5}) for
the $A_{jk}^{(0)}$ with parameters $\lambda _{\mathrm{th}}=4\sigma $,
$kT=5\hbar \gamma $. There we see clearly that the condition (\ref{5.3})
is violated for short times. It follows that for such times there must be
negative eigenvalues; the density matrix is not positive-definite. We
gain some insight into this phenomenon if we use the expressions
(\ref{3.5}) to expand
\begin{equation}
\mathrm{Tr}\{\rho ^{2}\}\cong 1+(1-4\frac{\sigma ^{2}}{\lambda
_{\mathrm{th} }^{2}})\gamma t+\cdots .  \label{5.10}
\end{equation} From this we see that $\mathrm{Tr}\{\rho ^{2}\}$ will
increase above unity whenever $\lambda _{\mathrm{th}}>2\sigma $. The
failure of positivity occurs for a sharply peaked wave packet.

To see explicitly this failure of positivity, it will be sufficient to
take
$ x_{0}=0$. We then form the expectation of the density matrix with a wave
function $\psi (x)\propto d\phi (x,0^{+})/dx$, where $\phi (x,0^{+})$ is
the state immediately after the coupling to the bath (given in Eq.
(\ref{3.12}) with $x_{0}=0$). Explicitly,
\begin{equation}
\psi (x)=\frac{x}{\sigma (2\pi \sigma ^{2})^{1/4}}\exp
\{-(\frac{1}{4\sigma ^{2}}+i\frac{m\gamma }{2\hbar })x^{2}\}. 
\label{5.12}
\end{equation} With this,
\begin{eqnarray}
\left\langle \psi ,\rho (t)\psi \right\rangle &=&\int_{-\infty }^{\infty
}dx\int_{-\infty }^{\infty }dx^{\prime }\psi ^{\ast }(x)\left\langle
x\left\vert \rho (t)\right\vert x^{\prime }\right\rangle \psi (x^{\prime
}) 
\nonumber \\ &=&\frac{-2[1-\frac{4}{\hbar
^{2}}(A_{11}^{(0)}A_{22}^{(0)}-A_{12}^{(0)2})]}{ [(1+\frac{A_{11}}{\sigma
^{2}})(1+\frac{4\sigma
^{2}(A_{11}^{(0)}A_{22}^{(0)}-A_{12}^{(0)2})}{A_{11}})+\frac{4\sigma
^{2}(A_{12}^{(0)}+m\gamma A_{11}^{(0)})^{2}}{\hbar ^{2}A_{11}}]^{3/2}}.
\label{5.13}
\end{eqnarray} This expectation will be negative whenever $
4(A_{11}^{(0)}A_{22}^{(0)}-A_{12}^{(0)2})/\hbar ^{2}<1$, that is exactly
for those times when $\mathrm{Tr}\{\rho ^{2}\}$ given by (\ref{5.9}) is
greater than one.

We have seen that general solution of the exact master equation results
in a density matrix that can have negative eigenvalues. How can it be
that an exact result can lead to such an unphysical consequence? The
answer, of course, is that we been able to give a non-trivial meaning to
the solution only in the high temperature limit, which involves a serious
approximation (neglect of zero-point oscillations). It is this
approximate result that fails the positivity test.

\subsection{Wave packet pair}

For the wave packet pair the Wigner characteristic function at time $t$ is
given by (\ref{4.3}). With this the formula (\ref{5.4}) for the matrix
element of the density operator becomes a standard Gaussian integral. The
result is
\begin{eqnarray}
\left\langle x\left\vert \rho (t)\right\vert x^{\prime }\right\rangle &=&
\frac{1}{2(1+e^{-d^{2}/8\sigma ^{2}})}  \nonumber \\ &&\times \lbrack \exp
\{iL\frac{d}{2}(x-x^{\prime })\}R(x-m\dot{G}\frac{d}{2} ,x^{\prime
}-m\dot{G}\frac{d}{2})  \nonumber \\ &&\exp \{-iL\frac{d}{2}(x-x^{\prime
})\}R(x+\frac{m\dot{G}d}{2},x^{\prime }+
\frac{m\dot{G}d}{2})  \nonumber \\ &&+e^{-A^{(0)}}\exp
\{iM\frac{d}{2}(x+x^{\prime })\}R(x-K\frac{d}{2} ,x^{\prime
}+K\frac{d}{2})  \nonumber \\ &&+e^{-A^{(0)}}\exp
\{-iM\frac{d}{2}(x+x^{\prime })\}R(x+K\frac{d}{2} ,x^{\prime
}-K\frac{d}{2})],  \label{5.13a}
\end{eqnarray} where $R(x,x^{\prime };t)$ is is the matrix element
(\ref{5.8b}) corresponding to a single wave packet initially centered at
the origin and to shorten the expressions we have introduced
\begin{equation} K=\frac{\hbar
^{2}(m\dot{G}A_{11}^{(0)}-GA_{12}^{(0)})}{4\sigma
^{2}(A_{11}^{(0)}A_{22}^{(0)}-A_{12}^{(0)2})},\quad
L=\frac{m^{2}\ddot{G}}{
\hbar }.,\quad M=\frac{\hbar (m\dot{G}A_{12}^{(0)}-GA_{22}^{(0)})}{4\sigma
^{2}(A_{11}^{(0)}A_{22}^{(0)}-A_{12}^{(0)2})}.  \label{5.13b}
\end{equation} Viewed in the $xx^{\prime }$ plane, the density matrix
element (\ref{5.13a}) will shows four peaks exactly of the form
(\ref{5.8a}) of single wave packets; two diagonal peaks centered at
$x=x^{\prime }=\pm m\dot{G}d/2$ \ and two off-diagonal peaks centered at
$x=-x^{\prime }=\pm Kd/2$. The off-diagonal peaks are multiplied by the
same factor $e^{-A^{(0)}(t)}$ that multiplies the interference term in
the Wigner function (\ref{4.14}). The result is that under the condition
$d\gg \lambda _{\mathrm{th}}$, while the peaks slowly broaden and drift
toward the origin, the off diagonal peaks rapidly shrink in amplitude.
Thus, the two descriptions of the disappearance of the interference term,
that in terms of the Wigner function and that in terms of the density
matrix element, are entirely equivalent. In this connection note that the
diagonal density matrix element is the probability distribution
(\ref{4.10}). The description in terms of the Wigner function has the
advantage of being real, so somewhat simpler. Neither the Wigner
distribution in phase space nor the density matrix element in $xx^{\prime
}$ space is directly observable, so either description is theoretical.

\section{Conclusion}

We have exhibited a number of significant failures of the exact master
equation: Due to the irreparable divergence the solution strictly does not
exist except in the high temperature limit. Even in that approximation the
solution breaks translational invariance. Finally, the solution
corresponds to a density matrix that can have negative expected values.
But we must remember that the derivation of the equation as well as its
solution are formally correct. Rather, the failures can all be traced
back to the assumption of an unentangeled initial state. The picture of a
particle suddenly clamped to a bath corresponds to no physically
realizable operation. Even when the initial particle temperature is
adjusted to that of the bath it receives a violent squeeze toward the
origin. From this follows the "unphysical" consequences: correctly
described by the exact master equation. Unfortunately, the final
assessment is that the exact master equation is of \ very limited utility.

In fact, in an exact description the particle must be coupled to the bath
(entangled) at all times. Think of the example of the blackbody radiation
field, which is coupled to the particle through its charge that can in no
way be switched on or off. In physical applications the coupling to the
radiation field is sufficiently weak that a description in terms of the
familiar weak coupling master equation (of Lindblad form) is appropriate,
but the effects of renormalization and zero-point oscillations are
nevertheless present at all times.

\newpage

\begin{figure}
\includegraphics[width=4.5in]{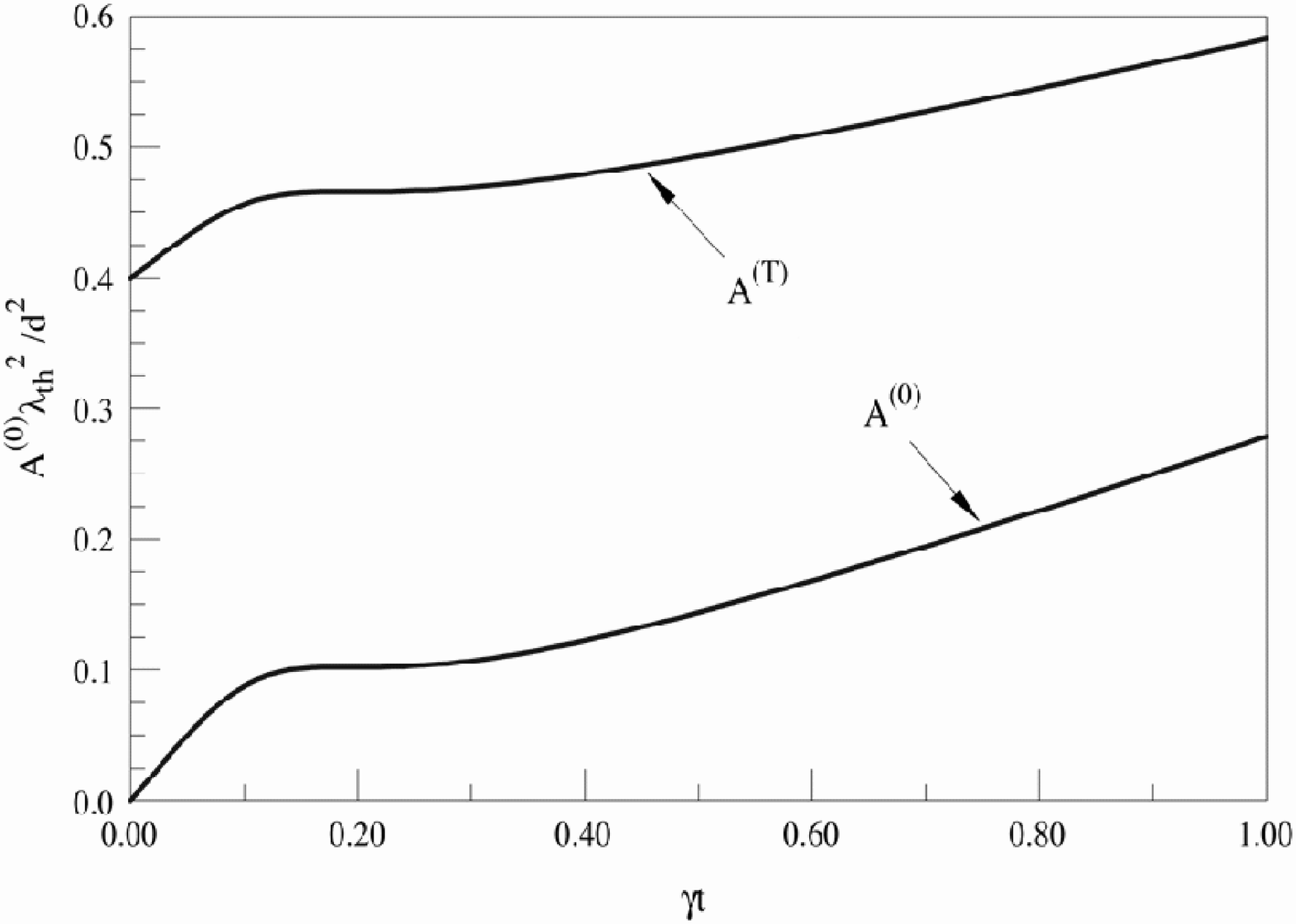}
\caption{The functions $A^{(0)}(t)$ and
$ A^{(T)}(t)$ multiplied by $\protect\lambda _{\mathrm{th}}^{2}/d^{2}$
plotted vs. $\protect\gamma t$. The parameters chosen are
$\protect\lambda _{\mathrm{ th}}=4\protect\sigma $ and $kT=5\hbar
\protect\gamma $. }
\label{lufig1}
\end{figure}

\begin{figure}
\includegraphics[width=4.5in]{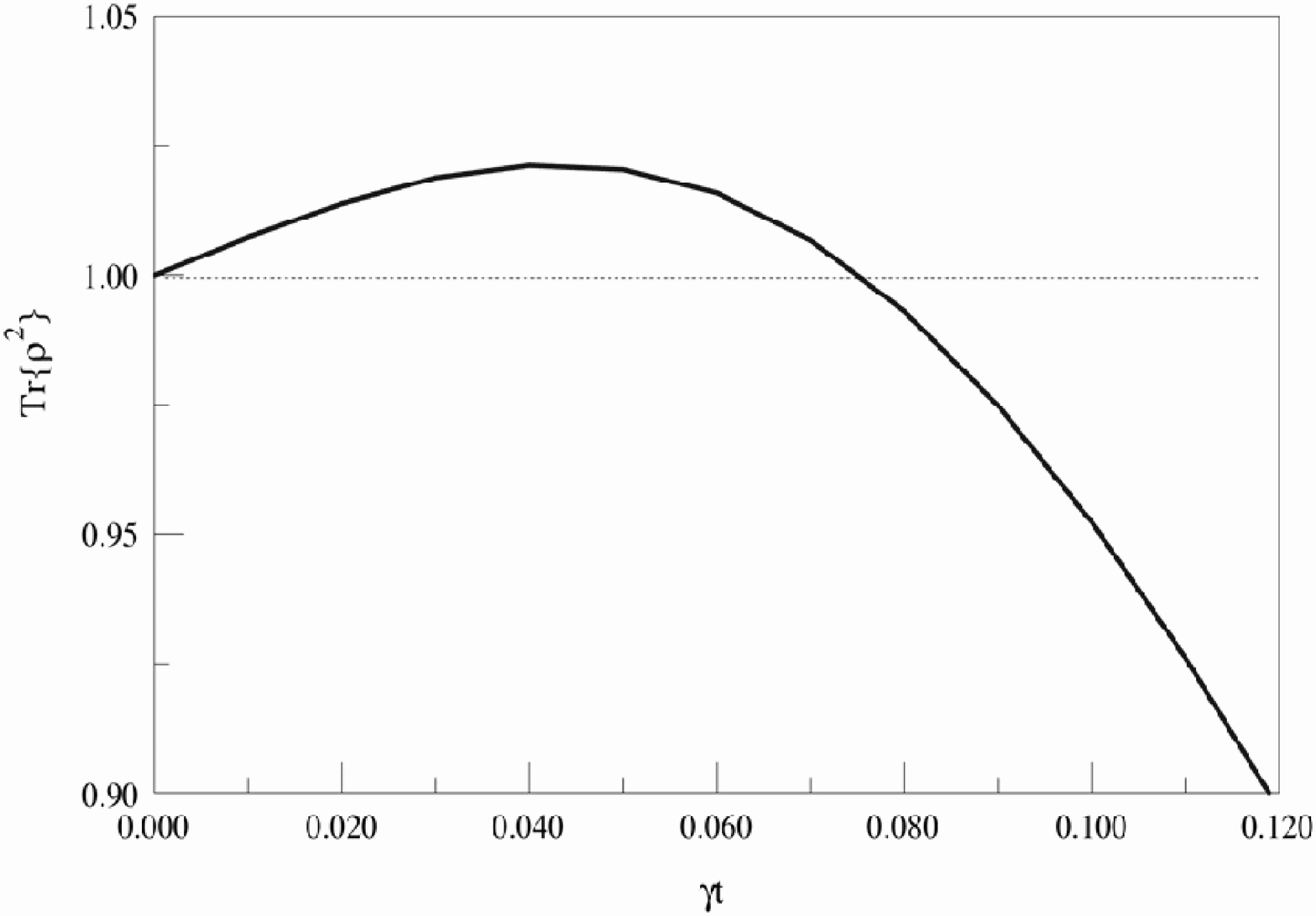}
\caption{$\mathrm{Tr}\{\protect\rho 
^{2}\} $ for an initial minimum uncertainty wave packet. The parameters
chosen are $\protect\lambda _{\mathrm{th}}=4\protect\sigma ,kT=5\hbar 
\protect\gamma $, the same as for Fig. 1. Note that at short times
$\mathrm{ Tr}\{\protect\rho
^{2}\}>1.$}
\label{lufig2}
\end{figure}


\newpage


\begin{thebibliography}{00}
\bibitem{ford01d} G. W. Ford and R. F. O'Connell, Phys. Rev. D {\bf{64}}
105020 (2001).

\bibitem{ford3} G. W. Ford and M. Kac, J. Stat. Phys. {\bf{45}}, 803
(1987).

\bibitem{vanhove52} L. van Hove, Physica {\bf{18}}, 145 (1952).

\bibitem{araki82} H. Araki and S. Yamagami, Publ. Res. Inst. Math. Sci. 
{\bf{18}}, 283 (1982).

\bibitem{efimov94} G. V. Efimov and W. von Waldenfels, Annals of Physics 
{\bf{233}}, 182 (1994).

\bibitem{lindblad} G. Lindblad, Commun. Math. Phys. {\bf{48}}, 119
(1976).

\bibitem{ford01a} G. W. Ford J. T. Lewis and R. F. O'Connell, Phys. Rev.
A 
{\bf{64}}, 032101 (2001).

\bibitem{garrett97} G. A. Garrrett, A. G. Rojo, A. K. Sood, J. F.
Whittaker and R. Merlin, Science {\bf{275}}, 1638 (1997).

\bibitem{ford02} G. W. Ford and R. F. O'Connell, American Journal of
Physics 
{\bf{70}}, 319 (2002).

\bibitem{murakami03} M. Murakami, G. W. Ford and R. F. O'Connell, Laser
Physics {\bf{13}}, 180 (2003).
\end{thebibliography}
\end{document}